\documentclass[JCPM]{iopart}
\usepackage{listings}
\usepackage{footmisc}
\usepackage{enumerate}
\usepackage{latexsym}
\usepackage{braket}
\usepackage{graphicx}
\usepackage[caption=false]{subfig}
\usepackage[colorlinks=True,linkcolor=red,citecolor=blue,urlcolor=blue]{hyperref}
\usepackage{blkarray}
\usepackage{array}
\usepackage{color}
\usepackage[normalem]{ulem}

\definecolor{orange}{rgb}{1,0.5,0}

\def\g{\Gamma}
\def\k{\vec{k}}

\newcommand\scalemath[2]{\scalebox{#1}{\mbox{\ensuremath{\displaystyle #2}}}}

\makeatletter
\DeclareRobustCommand{\element}[1]{\@element#1\@nil}
\def\@element#1#2\@nil{%
	#1%
	\if\relax#2\relax\else\MakeLowercase{#2}\fi}
\makeatother

\newcolumntype{L}{>{$}l<{$}}

\begin{document}
	\title{Spectral and optical properties of Ag${}_3$Au(Se${}_2$,Te${}_2$) and dark matter detection}
	\author{M.-\'{A}. S\'{a}nchez-Mart\'{i}nez~\footnote{\label{contrib} These authors contributed equally.}} 
	\address{Univ. Grenoble Alpes, CNRS, Grenoble INP, Institut N\'{e}el, 38000 Grenoble, France}
	\author{I. Robredo~\textcolor{red}{\footref{contrib}}}
	\address{Donostia International Physics Center, 20018 Donostia-San Sebastian, Spain}
    \address{Department of Condensed Matter Physics, University of the Basque Country UPV/EHU, Apartado 644, 48080 Bilbao, Spain}
	\author{A. Bidauzarraga}
	\address{Department of Condensed Matter Physics, University of the Basque Country UPV/EHU, Apartado 644, 48080 Bilbao, Spain}
	\author{A. Bergara}
	\address{Department of Condensed Matter Physics, University of the Basque Country UPV/EHU, Apartado 644, 48080 Bilbao, Spain}
	\address{Donostia International Physics Center, 20018 Donostia-San Sebastian, Spain}
	\address{Centro de Física de Materiales CFM, Centro Mixto CSIC-UPV/EHU, 20018 Donostia, Spain.}
	\author{F. de Juan}
	\address{Donostia International Physics Center, 20018 Donostia-San Sebastian, Spain}
    \address{IKERBASQUE, Basque Foundation for Science, Maria Diaz de Haro 3, 48013 Bilbao, Spain}
	\author{A. G. Grushin}
	\address{Univ. Grenoble Alpes, CNRS, Grenoble INP, Institut N\'eel, 38000 Grenoble, France}
	\author{M. G. Vergniory}
	\address{Donostia International Physics Center, 20018 Donostia-San Sebastian, Spain}
	\address{IKERBASQUE, Basque Foundation for Science, Maria Diaz de Haro 3, 48013 Bilbao, Spain}
	\date{\today}

\begin{abstract}
    In this work we study the electronic structure of Ag${}_3$AuSe${}_2$ and Ag${}_3$AuTe${}_2$, two chiral insulators whose gap can be tuned through small changes in the lattice parameter by applying hydrostatic pressure or choosing different growth protocols.  Based on first principles calculations we compute their band structure for different values of the lattice parameters and show that while Ag${}_3$AuSe${}_2$ retains its direct narrow gap at the $\Gamma$ point, Ag${}_3$AuTe${}_2$ can turn into a metal. Focusing on Ag${}_3$AuSe${}_2$ we derive a low energy model around $\Gamma$ using group theory, which we use to calculate the optical conductivity for different values of the lattice constant.  We discuss our results in the context of detection of light dark matter particles, which have masses of the order of a $k$eV, and conclude that Ag${}_3$AuSe${}_2$ satisfies three important requirements for a suitable detector: small Fermi velocities, $m$eV band gap and low photon screening. Our work motivates the growth of high-quality and large samples of Ag${}_3$AuSe${}_2$ to be used as target materials in dark matter detectors.
\end{abstract}	
\maketitle

\section{Introduction}

Narrow gap semiconductors belong to a particular branch of the semiconductor family, those with a narrow forbidden energy window, the gap, between valence and conduction bands. They were first applied as infrared detectors, with Hg${}_{1-x}$Cd${}_x$Te \cite{LAWSON1959325} as a representative example, due to its largely tunable band gap around infrared frequencies.

A particularly recent and promising application of narrow gap semiconductors is the direct detection of light dark matter particles~\cite{DMfito,Lin19,Geilhufe18}, an approach that complements those based on superconductors \cite{Hochberg16,Hochberg2016b,Hochberg16c} and phonons in polar materials~\cite{KNAPEN2018386,Griffin18}. These proposals are tailored to detect dark matter particles with $k$eV masses, which requires materials with gaps in the $m$eV range~\cite{Lin19} to match the expected energy deposition. To increase their sensitivity, narrow gap semiconductors must satisfy additional requirements, such as small Fermi velocities compared to the maximum dark matter velocities, or practical viability in terms of cost and purity~\cite{DMfito,Geilhufe18}. 

In this context, it is possible to achieve a highly tunable gap by changing the lattice parameter, either by applying hydrostatic pressure or by changing the method used to grow the crystal. This has the effect of bringing the ions closer (or farther) from each other, translating into small changes in the band structure. With ab-initio techniques, it is possible to predict the effect of small changes of the lattice parameter in the bands, in particular the magnitude of the gap.

In this work we study the effect that small changes in the lattice parameter have on the properties of Ag${}_3$AuSe${}_2$ and Ag${}_3$AuTe${}_2$, two chiral narrow gap semiconductors. We perform a theoretical analysis of the band structure for both compounds, which share spectral properties, and we focus on Ag${}_3$AuSe${}_2$ to calculate its optical conductivity using an effective model for the bands near the Fermi level. We conclude by discussing the suitability of this material to detect light dark matter particles.

\section{Analysis based on Density Functional Theory}
\begin{figure}[t]
	\centering
	\includegraphics[width=1.\linewidth]{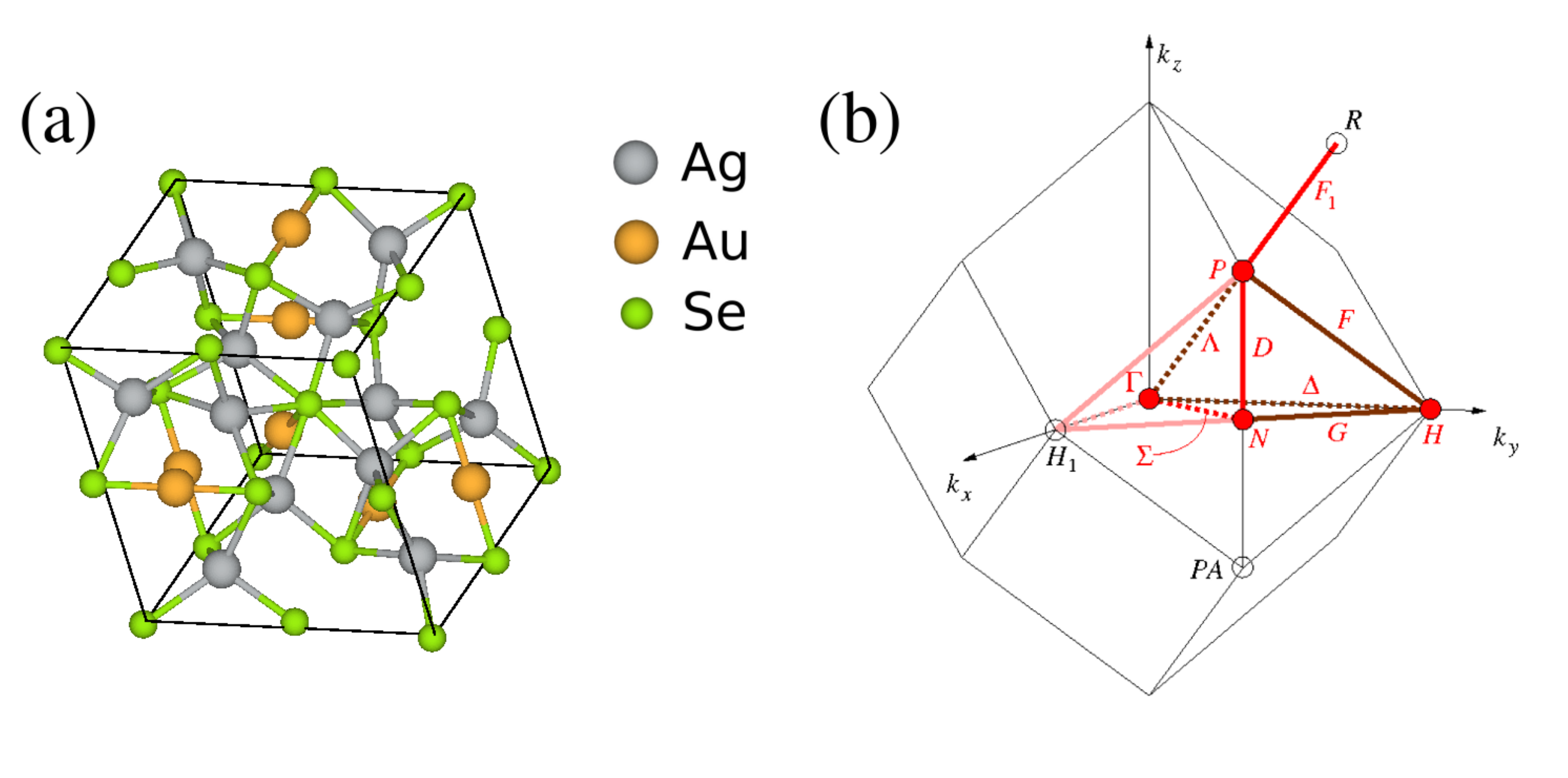}
		\caption{Crystalline structure (a) and Brillouin zone~\cite{BCS2,BCS3} (b) of Ag${}_3$AuSe${}_2$.}
		\label{fig:structure}
\end{figure}

Ag${}_3$AuSe${}_2$ and Ag${}_3$AuTe${}_2$ crystallize in the Sohncke space group $I4_132$ (space group 214, see Fig.~\ref{fig:structure}), and are predicted to hold high order nodal points in their band structure~\cite{Bradlynaaf5037}. Both have been diagnosed as trivial insulators in a recent work~\cite{TQC,High-Quality,TQCweb}. Space group $I4_132$(214) is a chiral non-symmorphic space group; it is not centro-symmetric or mirror-symmetric and some of the symmetry operations contain non-integer translations~\cite{bradley}.

We have used Density Functional Theory (DFT)~\cite{DFT1,DFT2} as implemented in the Vienna Ab initio Simulation Package (VASP)~\cite{VASP1,VASP2} to perform band structure calculations. To account for the interaction between ion cores and valence electrons we used the projector augmented-wave method (PAW)~\cite{PAW}. We use the generalized gradient approximation (GGA) for the exchange-correlation potential with the Perdew-Burke-Ernkzerhof functional for solids parameterization~\cite{PBE} and consider the spin-orbit coupling (SOC) interaction as implemented within the second variation method~\cite{SOC}. We have used a Monkhorst-Pack k-point grid of ($7\times7\times7$) for reciprocal space integration and a 500 eV energy cutoff of the plane-wave expansion.

\begin{figure}[t]
	\centering
	\includegraphics[width=0.45\linewidth]{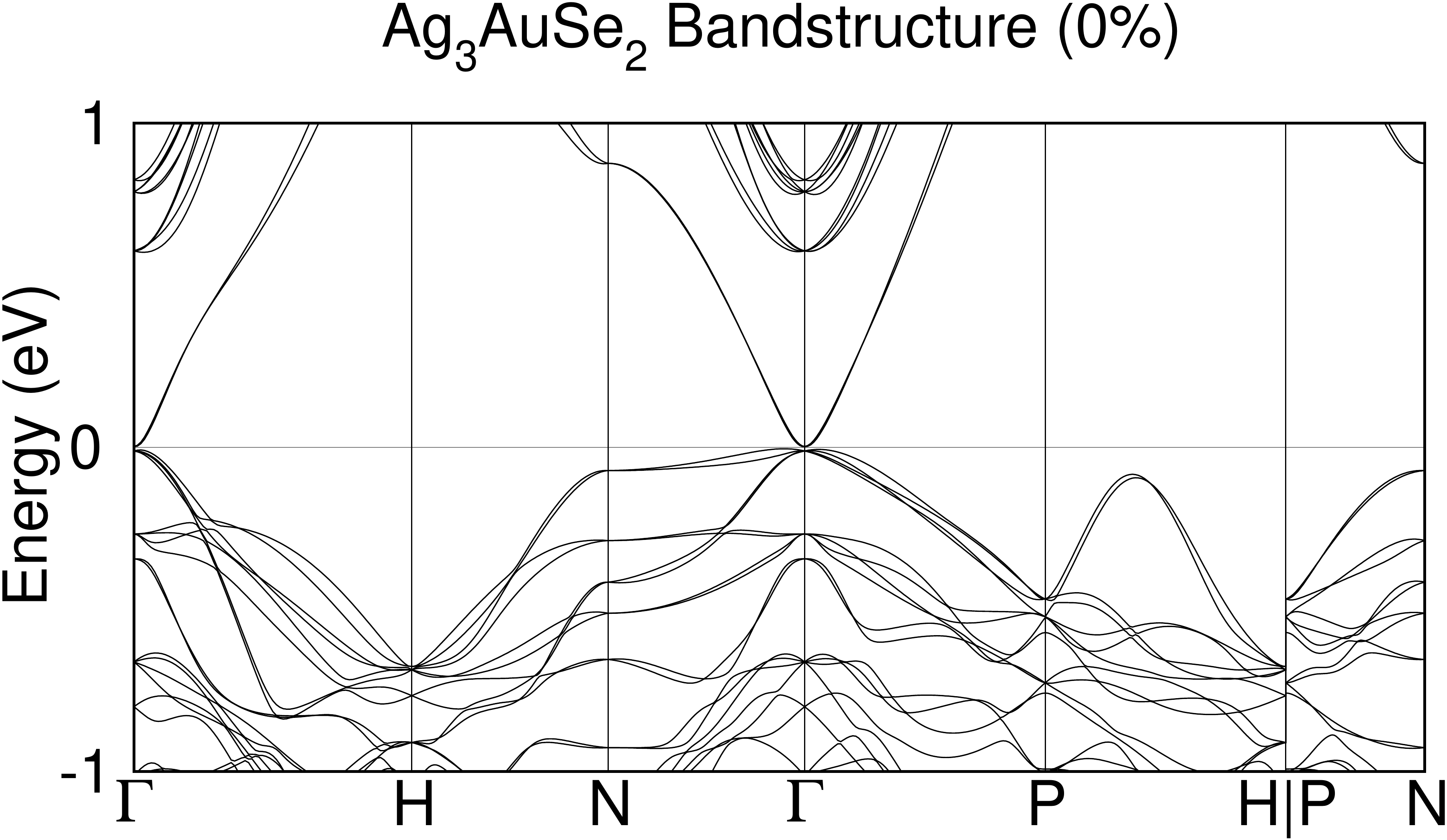}
	\includegraphics[width=0.45\linewidth]{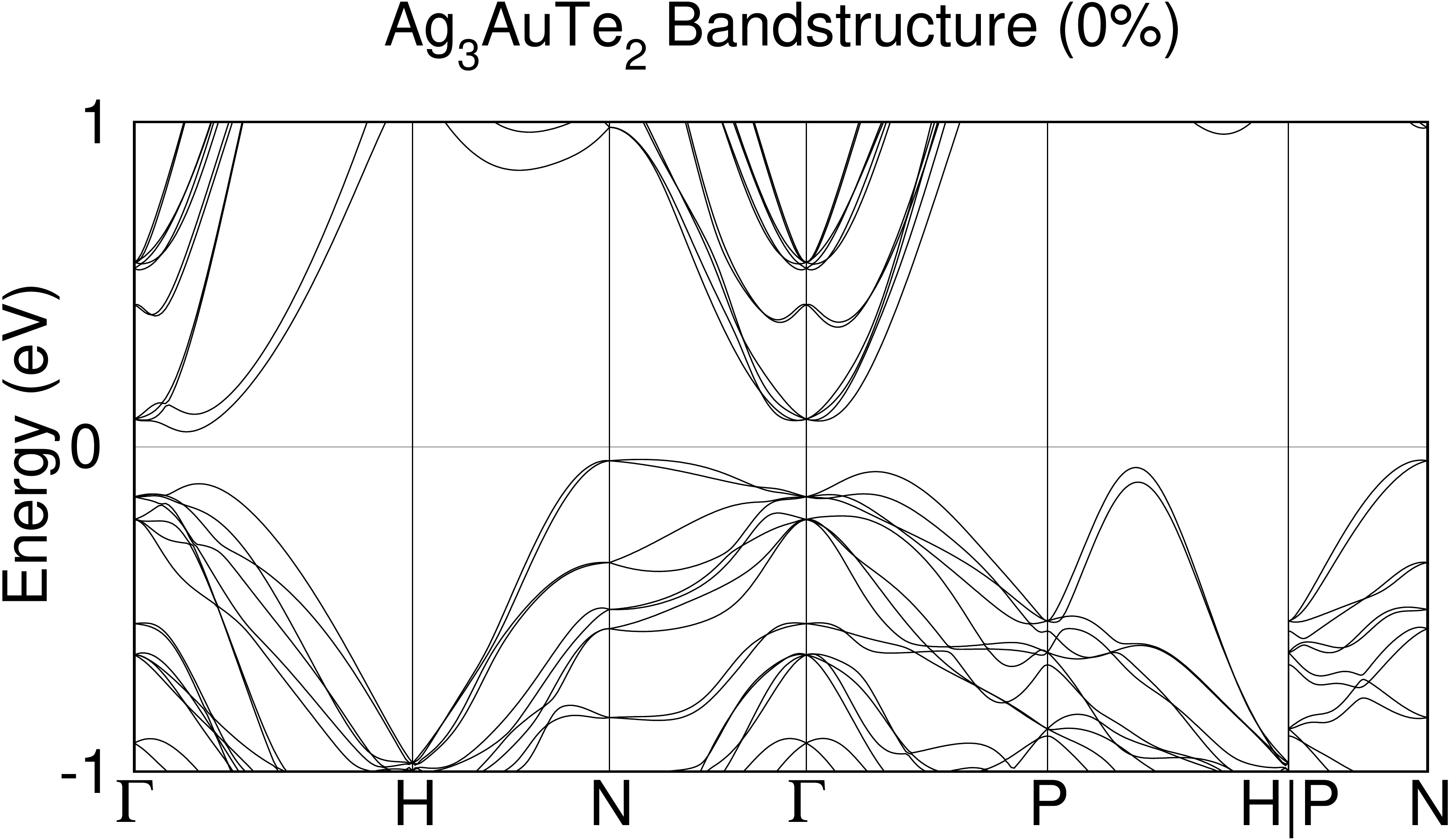}
	\includegraphics[width=0.45\linewidth]{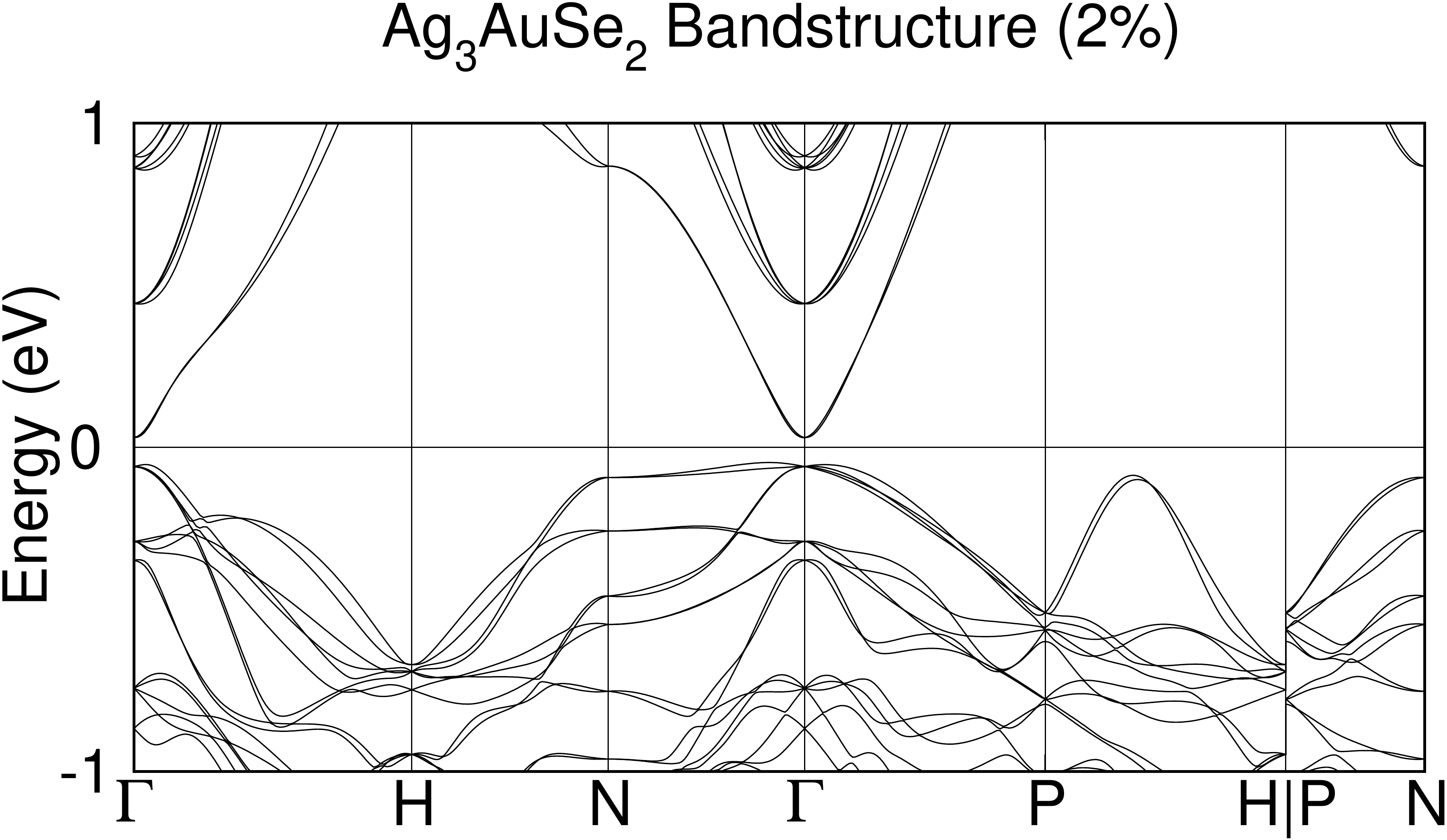}
	\includegraphics[width=0.45\linewidth]{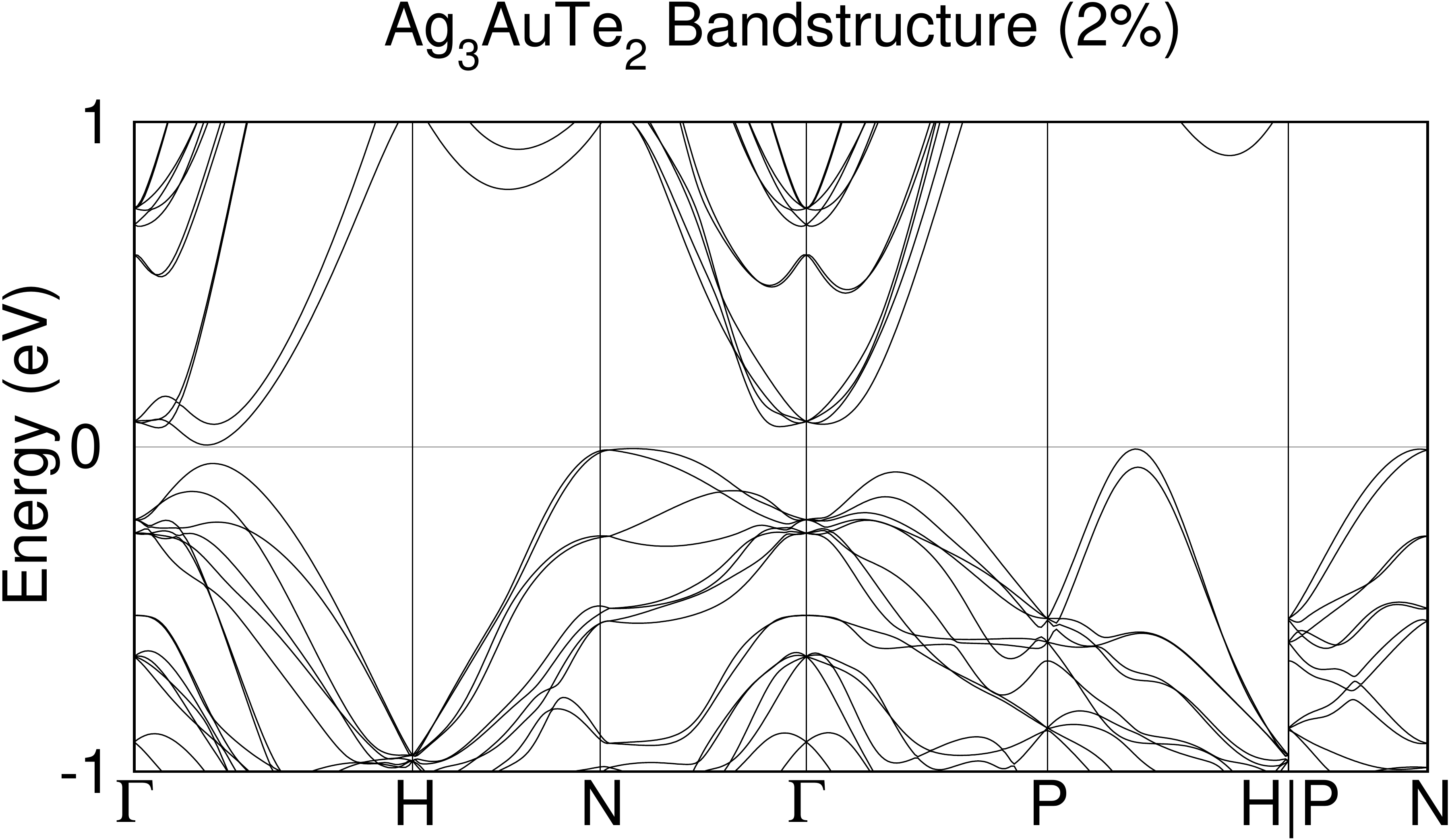}
	\includegraphics[width=0.45\linewidth]{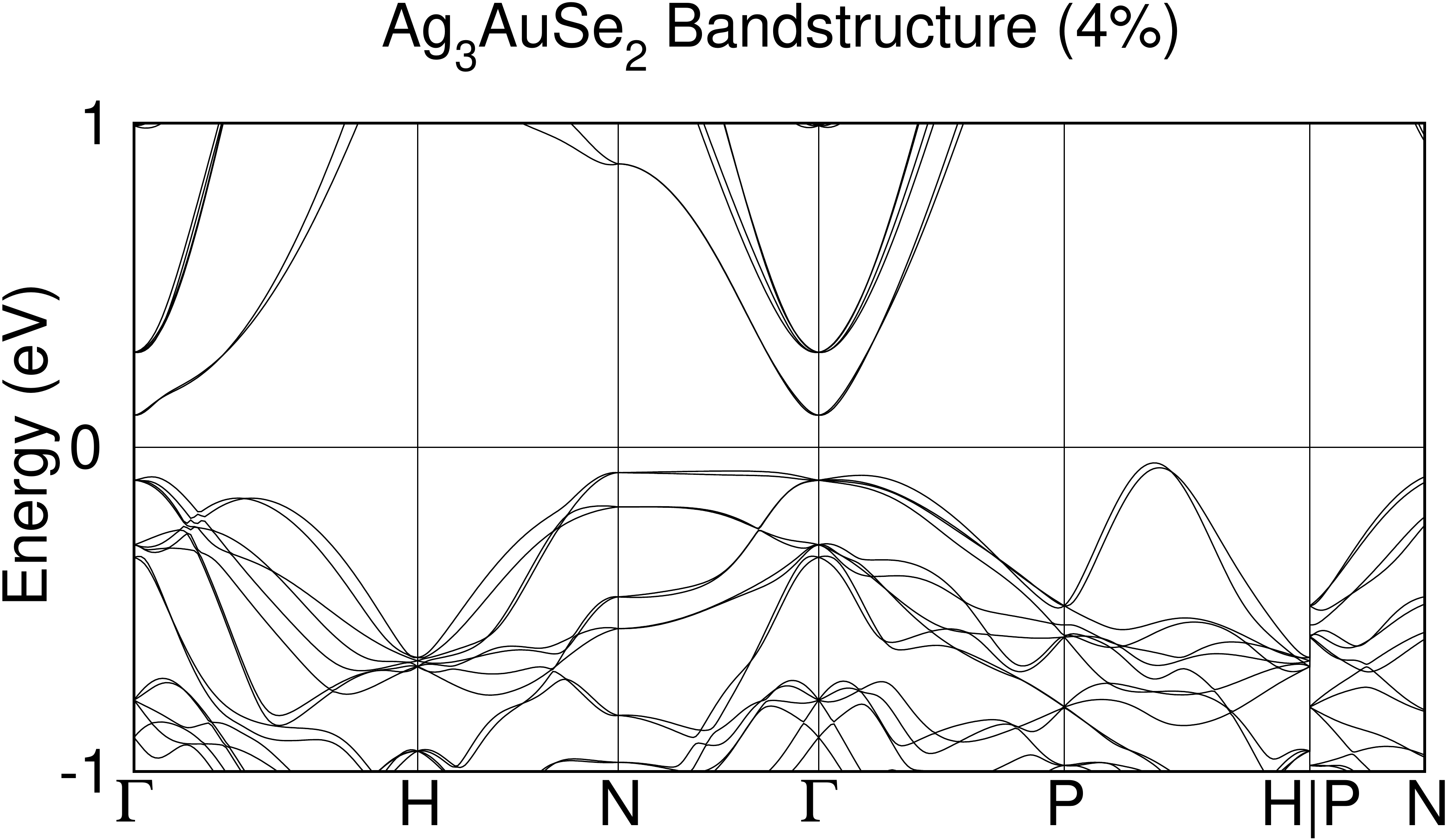}
	\includegraphics[width=0.45\linewidth]{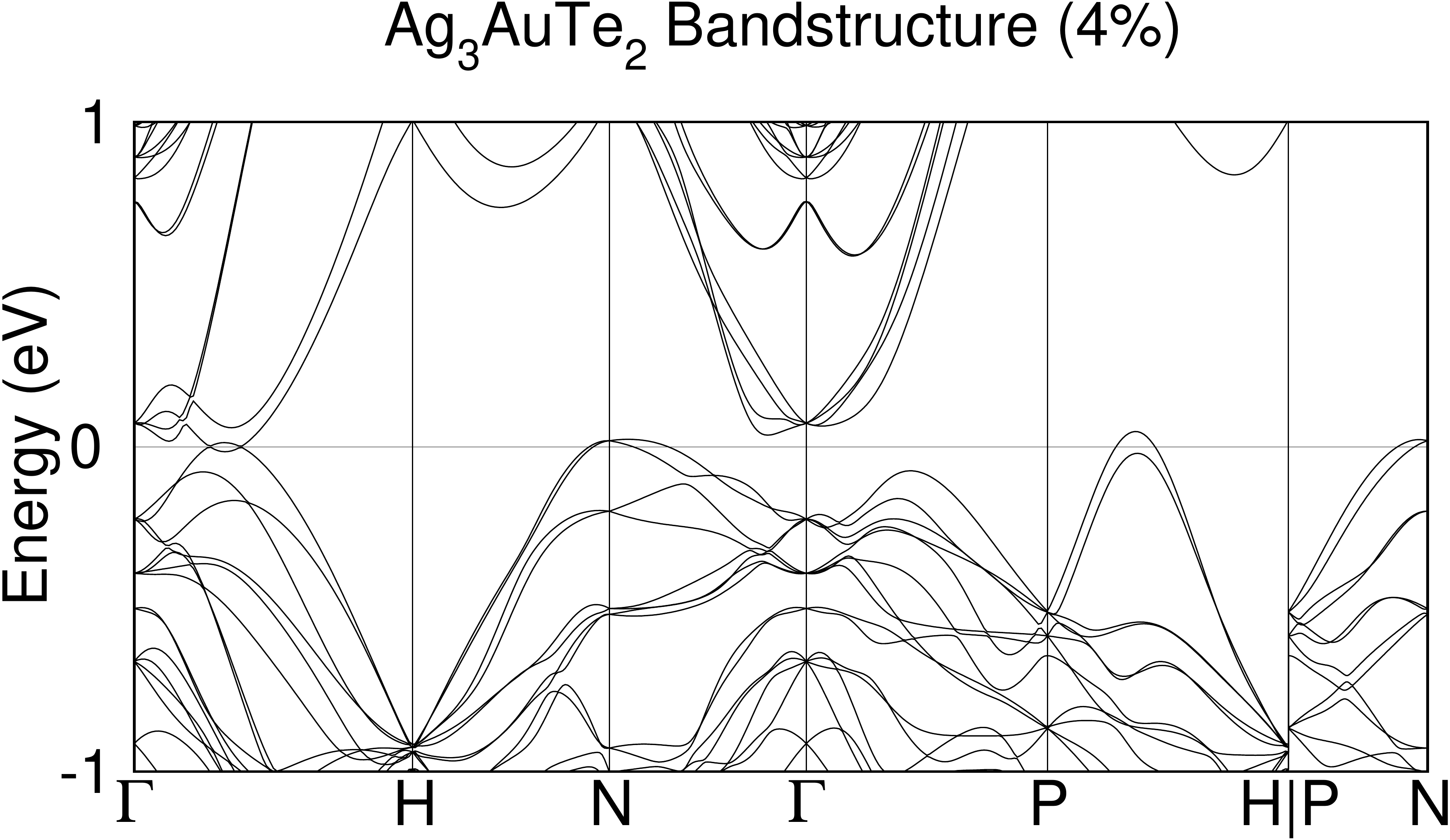}
		\caption{Bandstructures of Ag${}_3$AuSe${}_2$ and Ag${}_3$AuTe${}_2$ for different choices of the parameters. By reducing the lattice parameter the gap widens for the Se compound while not changing the overall band dispersion. The Te compound, however, becomes metallic.}
		\label{fig:CompressBands}
\end{figure}

We have computed the band structures for both compounds, Ag${}_3$AuSe${}_2$ and Ag${}_3$AuTe${}_2$, for different values of the lattice parameters along high-symmetry points of the Brillouin zone. Fig.~\ref{fig:CompressBands} shows that the gap of Ag${}_3$AuSe${}_2$ increases from 13.8 meV to 60 meV for a compression of 2\%,  and reaches 200 meV for a compression of 4\% at $\g$, keeping it as a global direct gap.
This implies that the interband optical conductivity and the suitability as a dark matter detector, both relaying on the interband transitions across the gap~\cite{DMfito}, will be determined by the bands in the neighborhood of the $\g$ point for a wide range of frequencies. The Te-based compound on the other hand  is a semiconductor with small indirect gap originally, and the change in the lattice parameter transforms it into a metal. Therefore, among these two, Ag${}_3$AuSe${}_2$ is the most promising for detector applications, due to its small tunable direct band gap at $\g$. In the next section we construct an effective model around this high-symmetry point to further analyze its optical conductivity and potential as a dark matter detector. 

\begin{figure}[t]
	\centering
	\includegraphics[width=1\linewidth]{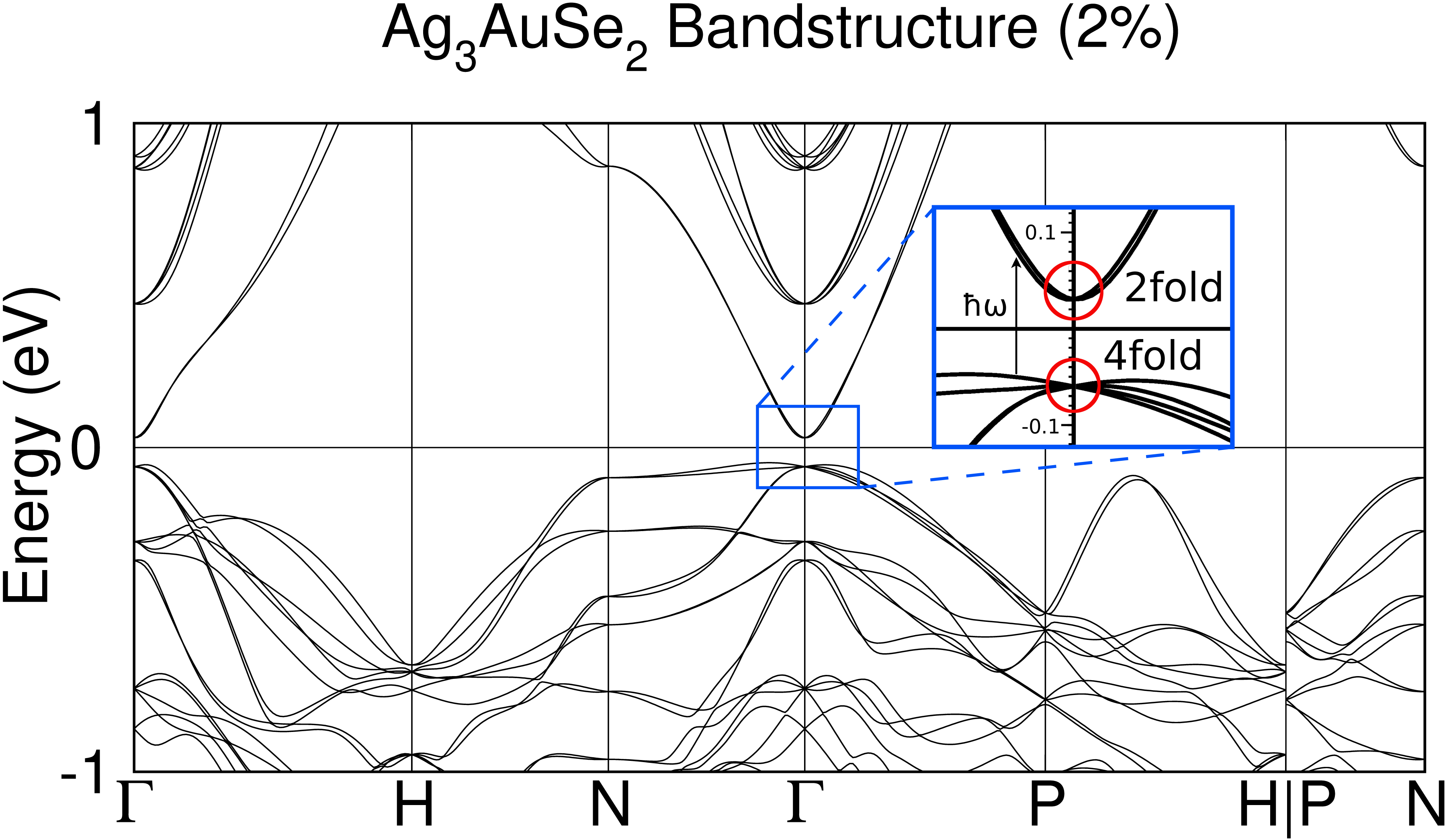}
		\caption{Band structure of Ag${}_3$AuSe${}_2$ with lattice parameter 2\% smaller than the original. Inset: the narrow gap at $\g$ lies between a twofold fermion (Weyl fermion) and a fourfold-fermion. The interband transition across the gap is the relevant to our optical conductivity, and is shown schematically with a vertical arrow. Each horizontal tick corresponds to 10 meV.}
		\label{fig:Bands}
\end{figure}

\section{Effective model}

In this section we use group theory to construct the most general low energy Hamiltonian allowed by symmetry for the four last valence and first two conduction bands. We also allow terms that couple the valence and conduction band sub-spaces, resulting in a six-band model. To determine the free parameters of the effective Hamiltonian we fit its spectrum to the ab-initio bands found in the previous section for different values of the lattice constant.

\subsection{How to build effective models}

In general, an effective Hamiltonian, which will describe the crossing of $n$ bands, can be written as

\begin{equation}
\label{eq1}
H(\vec{k})=\Psi^*_i H_{i,j}(\vec{k})\Psi_j=\sum_{i,j}^n\Psi^*_i\Psi_jH_{i,j}(\vec{k})=\sum_a \left(\Psi^*\Psi\right)^a H_a,
\end{equation}
where $\Psi_i$ are the Bloch states of the Hamiltonian and $\left(\Psi^*\Psi\right)^a$ are the bilinears, that can be written as 
\begin{equation}
\left(\Psi^*_i\Psi_j\right)^a=\lambda^a_{ij},
\end{equation}
where $a\in\{0,1,2,...,n^2\}$ labels the bilinear and $\lambda^a$ are complex Hermitian matrices that form a basis under which the Hamiltonian can be expanded.

In the following we will develop the low energy Hamiltonian for Ag${}_3$AuSe${}_2$. Different values of the model parameters correspond to different values of the lattice constant by which we take into account different values of the hydrostatic pressure. To compute the optical conductivity and asses the absorption of the material due to transitions across the gap we aim to construct the lowest order model that describes optical transitions across the minimal gap, which occurs at the $\g$ point. We focus on the first four valence bands below the Fermi level, forming a 4 dimensional energy crossing protected by symmetry (fourfold fermion), and the first two conduction bands above the Fermi level, described by a two-dimensional crossing also protected by symmetry (twofold fermion), as shown in the inset of Fig.~\ref{fig:Bands}.

To construct Hamiltonians invariant under the point group operations, we need to find combinations of bilinears and powers of momentum that transform under the trivial representation of the little group at $\g$, which is isomorphic to the Point Group $O$ (432). Powers of crystal momentum will transform under symmetry operations as a representation $\rho$. In particular, for this group the representation under which momentum transforms reads
\begin{eqnarray}
\label{eq:krep}
\rho(\k)=T_1(k_x,k_y,k_z)\oplus A_1(|\k|^2) \oplus 
T_2(k_yk_z,k_zk_x,k_xk_y)+O(k^3),
\end{eqnarray}
where $T_1,A_1$ and $T_2$ are irreducible representations of the Point Group $O$. Given the representation expansion of the momentum in Eq.~\ref{eq:krep}, we can form invariants from the combination of bilinears that transform under those same representations. The twofold band crossing is composed of two quadratic bands, and thus we consider second order terms in the momentum. For the fourfold band crossing and the coupling terms between the fourfold and twofold sub-spaces linear order in $\vec{k}$ will suffice. This will give us the lowest possible order Hamiltonian that contributes to the optical conductivity.

\subsection{Effective Hamiltonian for conduction bands}

In this case we need to find bilinears that transform under $T_1$, $A_1$ and $T_2$. In the basis of Pauli matrices, the bilinears transform as
\begin{equation}
\rho(\sigma_\mu)=A_1(\sigma_0)\oplus T_1(\vec{\sigma}).
\end{equation}
The absence of $T_2$ in the expansion implies that there is no symmetry-allowed term that can go with $T_2(k_yk_z,k_zk_x,k_xk_y)$. Thus, the effective Hamiltonian reads
\begin{equation}
H_{2band}=\alpha|\k|^2\sigma_0+v_F\k\cdot \vec{\sigma}=
\left(\begin{array}{cc}
\alpha|\k|^2 +v_Fk_z & v_F(k_x-ik_y) \\
v_F(k_x+ik_y) & \alpha|\k|^2 -v_Fk_z
\end{array}\right),
\end{equation}
%
%
where $\alpha$ and $v_F$ are the coefficients that will be fitted to our ab-initio results.

\subsection{Effective Hamiltonian for valence bands}

For the fourfold band crossing a basis of $4\times4$ Hermitian matrices is required. Since there are only  two sets of matrices that transform under the $T_1$ representation ($\vec{\lambda_1}$ and $\vec{\lambda_2}$), there are only two linear terms in momentum allowed in the Hamiltonian

\begin{equation}\label{eq:model4-1}
H_{4band}=v^1_F\k\cdot\vec{\lambda_1}+v^2_F\k\cdot\vec{\lambda_2}\\
\end{equation}
\begin{equation*}\label{eq:model4-2}
\fl
=\left(\scalemath{0.77}{
\begin{array}{cccc}
k_z v^1_F & (k_x-i k_y) v^1_F & \left(e^{-\frac{11 i \pi}{12}} k_x+e^{-\frac{i \pi}{12} } k_y\right) v^2_F & e^{-\frac{3 i \pi}{4} } k_z v^2_F \\
(k_x+i k_y) v^1_F & -k_z v^1_F & e^{\frac{3 i \pi }{4}} k_z v^2_F & \left(e^{\frac{7 i \pi }{12}} k_x+e^{\frac{5 i \pi }{12}} k_y\right) v^2_F \\
\left(e^{\frac{11 i \pi }{12}} k_x+e^{\frac{i \pi }{12}} k_y\right) v^2_F & e^{-\frac{3 i \pi}{4} } k_z v^2_F & -k_z v^1_F & (k_y-i k_x) v^1_F \\
e^{\frac{3 i \pi }{4}} k_z v^2_F & \left(e^{-\frac{7 i \pi}{12} } k_x+e^{-\frac{5 i \pi}{12} } k_y\right) v^2_F & (i k_x+k_y) v^1_F & k_z v^1_F \\
\end{array}}
\right).
\end{equation*}
\normalsize

\subsection{Coupling Hamiltonian}

So far we have computed the effective Hamiltonians for the twofold band and fourfold band crossings. Since these two Hilbert subspaces are orthogonal to each other, the matrix elements of the interband current operator, relevant to compute the optical conductivity, will vanish. To describe transitions accross the gap that connect the fourfold and twofold bands we have to add symmetry allowed terms that mix these two subspaces. There is only one allowed term at first order in momentum, which reads

\begin{equation}
\fl
H_{cp}=\delta\left(\scalemath{0.9}{
\begin{array}{cccc}
-k_x &  e^{\frac{i \pi }{3}} k_y+ e^{-\frac{5 i \pi}{6}} k_z & - e^{\frac{5 i \pi }{12}} k_y- e^{\frac{7 i \pi }{12}} k_z &  e^{\frac{i \pi }{4}} k_x \\
e^{\frac{i \pi }{3}} k_y- e^{-\frac{5 i \pi}{6} } k_z & k_x & - i e^{\frac{i \pi }{4}} k_x &  i e^{\frac{5 i \pi }{12}} k_y- i e^{\frac{7 i \pi }{12}} k_z \\
\end{array}}
\right).
\end{equation}

\subsection{Low energy Hamiltonian and parameters}
Combining the above results the full effective Hamiltonian reads
\begin{equation}
\label{eq:fullham}
H(\vec{k})=
\left(\begin{array}{cc}
H_{2band}+\Delta 1_{2\times2} & H_{cp} \\
H_{cp}^\dagger & H_{4band}-\Delta 1_{4\times4}
\end{array}\right),
\end{equation}
%
%
where we have added the parameter $\Delta$ that sets the gap between the twofold and fourfold crossings to be $2\Delta$.
We obtained the values of the parameters in the Hamiltonian Eq.~(\ref{eq:fullham}) by fitting ab-initio bands for three different choices of lattice parameters. Labelling the unperturbed lattice parameter as $a_0$~\cite{ICSD}, we define the lattice parameter under pressure as $a=pa_0$ for $p=0.99,0,98,0.97$, which amounts to compressions by 1\%, 2\%, and 3\%, respectively. We display the fitted values in Table \ref{tab:fitvals}, and the corresponding bands are plotted in Fig.~\ref{fig:Opt1} (a) and (b) for 1\% compression, and Fig.~\ref{fig:opt2} (a) and (b) for 2\% and 3\% compression, respectively.

\begin{table}[t]
\centering
	\begin{tabular}{c|c c c}
		& 1\% & 2\% & 3\% \\\hline
		$v_F^s$ & 0.164 & 0.141 & 0.105 \\
		$v_F^a$ & 0.228 & 0.211 & 0.178 \\
		$v_F$ & 0.390 & 0.398 & 0.371 \\
		$\alpha$ & 55.637 & 50.891 & 44.694 \\
		$\delta$ & 0.370 & 0.587 & 0.690 \\
		$\Delta$ & 0.009 & 0.030 & 0.056
	\end{tabular}
\caption{Fitted values for different choices of lattice parameter compression. The units are expressed in terms of $\hbar=c=1$ with  $\vec{k}$ in units of $2\pi/a$, where $a = p a_0$, $a_0$ is the unperturbed lattice constant and $p=0.99,0.98,0.97$ for $1\%$, $2\%$, $3\%$ hydrostatic pressure respectively.}
\label{tab:fitvals}
\end{table}

\section{Optical conductivity}

In this section we compute and discuss the interband optical conductivity of the six-band $\mathbf{k}\cdot\mathbf{p}$ Hamiltonian describing the band structure of Ag$_3$AuSe$_2$ near the $\Gamma$ point for three different hydrostatic pressures: $1\%$, $2\%$, and $3\%$ compression of the lattice parameter. In all three cases, the band structure exhibits a fourfold node below the Fermi level and a Weyl node above it, separated by a gap whose size will depend on the pressure.\\
The interband contribution to the conductivity tensor $\sigma^{\mu\nu}$, where $\mu,\nu=x,y,z$, can be calculated using standard linear
response theory as the real part of~\cite{Mahan}
\begin{equation}
\label{eqn:optcond}
\sigma_{\mu \nu}(\omega)=\frac{ie^2}{\omega V}\sum_{m\neq n}\frac{\bra{n} j_{\mu}\ket{m}\bra{m}j_{\nu}\ket{n}}{\epsilon_n-\epsilon_m+\hbar\omega+i\delta}\left(n_F(\epsilon_{n})-n_F(\epsilon_{m})\right),
\end{equation}
where $j_{\mu}=\frac{1}{\hbar}\partial_{k_{\mu}}H$ is the current operator defined by the Hamiltonian, $e$ is the charge of the electron, $V$ is the volume, $\ket{n}$ and $E_n$ are an eigenstate and the corresponding eigenvalue of the Hamiltonian, respectively, $\delta$ is an infinitesimal broadening; $\epsilon_n=E_n-\mu$, where $\mu$ is the chemical potential and $n_F$ is the Fermi function, which depends on $\epsilon_n$, $\mu$, and the inverse temperature $\beta=1/k_B T$ in units of the Boltzmann constant $k_B$. 

\subsection{Optical conductivity and relevant transitions for a given lattice parameter}
\begin{figure}[t]
	\centering
	\includegraphics[width=1\linewidth]{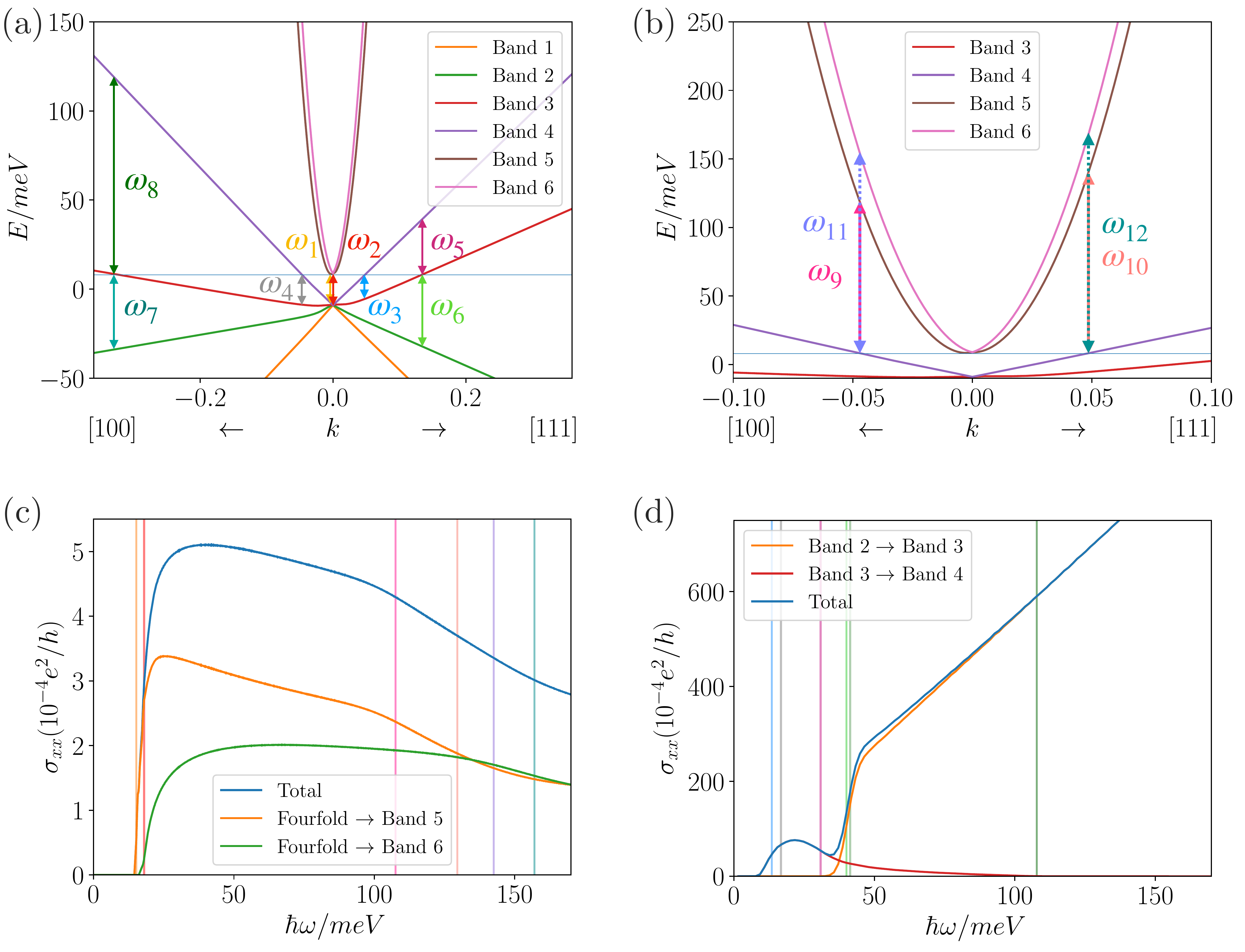}
		\caption{Optical conductivity of the low energy $\mathbf{k}\cdot\mathbf{p}$ model of Ag$_3$AuSe$_2$ under $1\%$ hydrostatic pressure. (a) and (b) show the same band structure within two different energy windows, together with the relevant frequencies for optical transitions (colored vertical arrows) and the chemical potential (blue horizontal line). The bands are labelled with numbers from $1$ to $6$ from bottom (orange band) to top (pink band). Figures (c) and (d) show the optical conductivity corresponding to valence to conduction band transitions (c) and interband transitions between the valence bands (d), with $\beta=1/k_B T=2\times 10^3$. }
		\label{fig:Opt1}
\end{figure}

As discussed in the previous section, for the three values of compression 1\%, 2\%, and 3\%, the bands are separated into a set of four bands, which we refer to as valence bands, that define the fourfold fermion below the chemical potential, and a set of two bands, which we refer to as the conduction bands, that define the Weyl fermion above the chemical potential. Note that while in in the full ab-initio bands there exists a full gap between the two sets of bands (see e.g. Fig.~\ref{fig:Bands}), in the low energy model the gap is not fully open.

To understand the relevant optical transitions and their contribution to the optical conductivity 
we first fix the lattice constant to a compression of $1\%$ by choosing the corresponding set of parameters (see Table \ref{tab:fitvals}). The band structure of the resulting $\mathbf{k}\cdot \mathbf{p}$ Hamiltonian is shown in Fig.~\ref{fig:Opt1}~(a) and (b) with two different scales for better readability, together with the relevant activation frequencies. The energies involved in the transitions between the fourfold fermion and the Weyl fermion are smaller than in the other two cases, which allows us to study the behavior of the optical conductivity with all possible activation frequencies in a smaller energy range.

The transitions between the six bands can be divided in two groups: the ones that only concern transitions between the valence bands, with activation frequencies $\omega_{i}$ with $i \in [3,8]$, and those corresponding to transitions connecting the valence and the conduction bands, with activation frequencies $\omega_i$ with $i=1,2$ and $i \in [9,12]$. Since in the real material the transitions between the fourfold valence bands will be Pauli blocked, we separate them from those connecting the valence and conduction bands, which are the only ones allowed in the real material. This justifies our choice of chemical potential: it is close to the Weyl node, shifting the activation frequencies between bands of the fourfold to higher frequencies.

In Fig.~\ref{fig:Opt1}~(c) and (d) we show the optical conductivity associated to both types of transitions, with the characteristic frequencies represented by vertical lines matching the colors in Fig.~\ref{fig:Opt1}~(a) and (b). The valence to conduction transitions are shown in Fig.~\ref{fig:Opt1}~(c), which shows a characteristic maximum that falls off as the frequency is increased. On the other hand, Fig.~\ref{fig:Opt1}~(d) shows the transitions between valence bands, which display a linear behaviour $\sigma \propto \omega$ characteristic of an asymmetric fourfold fermion~\cite{optcondmultifolds}. In both cases, the changes in the optical conductivity match well the activation frequencies expected from the band structure in Fig.~\ref{fig:Opt1}~(a) and (b). 

The optical conductivity of the six-band $\mathbf{k}\cdot\mathbf{p}$ Hamiltonian applies strictly to a narrow region in momentum-energy space yet Fig.~\ref{fig:Opt1} is enlightening to understand what to expect from the optical conductivity of the real Ag$_3$AuSe$_2$ and, in the next section, its usefulness to detect light dark matter.

\begin{figure}
	\includegraphics[width=1\linewidth]{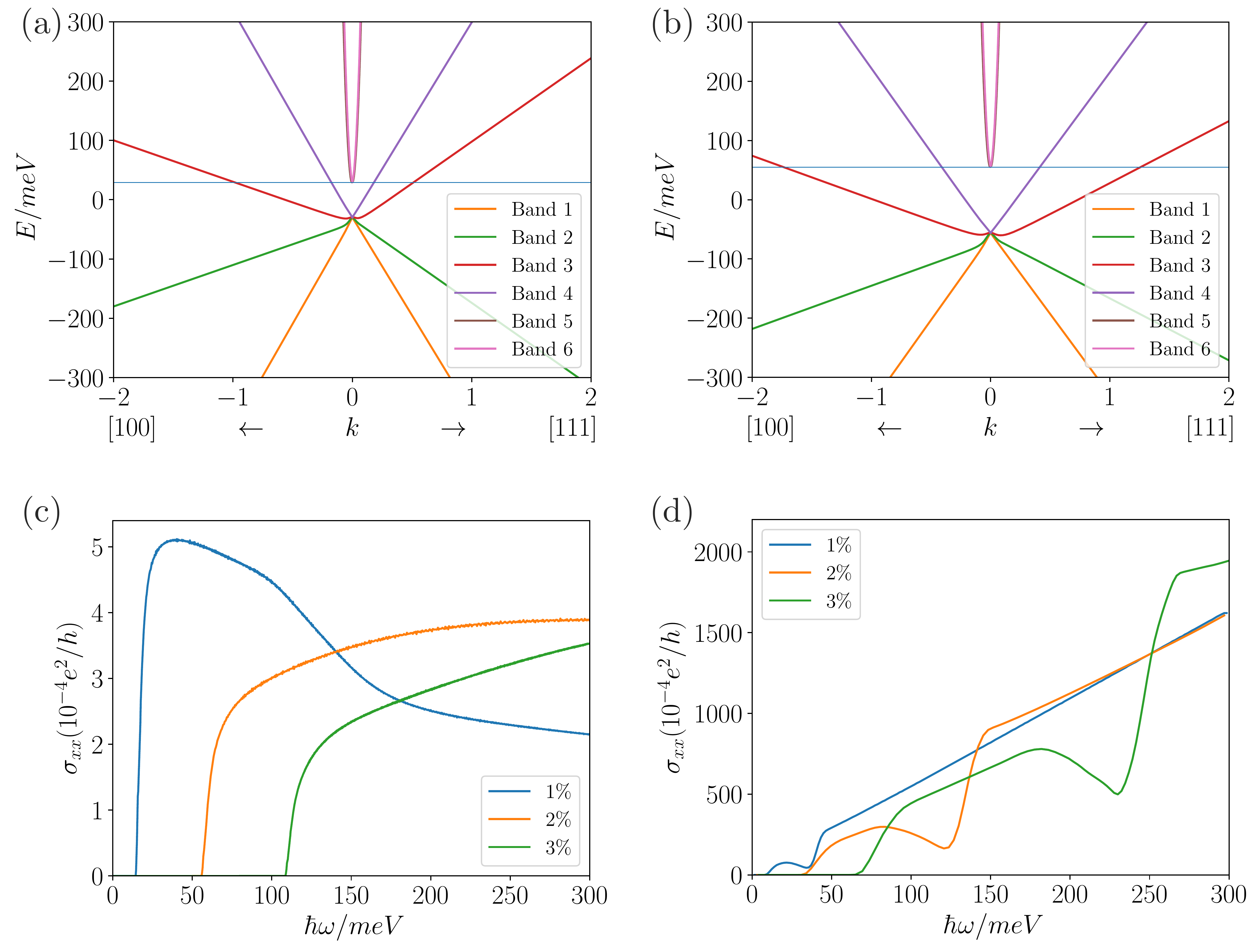}
		\caption{Optical conductivity of the low energy $\mathbf{k}\cdot\mathbf{p}$ model of Ag$_3$AuSe$_2$ under different hydrostatic pressures ($1-3\%$). (a) and (b) show the band structures for $2\%$ and $3\%$, respectively, with the same color coding as the bands for the $1\%$ case in Fig.~\ref{fig:Opt1}. (c) and (d) show the contribution to the optical conductivity arising from interband transitions between valence and conduction bands (c) and between valence bands (d), with $\beta=1/k_B T=2\times 10^3$.}
		\label{fig:opt2}
\end{figure}

\subsection{Optical conductivity for different lattice parameters}

As we increase the pressure to $2\%$ (Fig.~\ref{fig:opt2} (a)) and $3\%$ (Fig.~\ref{fig:opt2} (b)) the gap becomes larger, and the parameters describing both the fourfold and the Weyl fermion are modified (see Table \ref{tab:fitvals}). The Fermi velocity of the bands for each case, an important parameter for light dark matter detection~\cite{DMfito}, as well as the off-diagonal matrix elements connecting the fourfold and the Weyl fermions, change. Upon doing so, they modify the energy scales involved in the optical transitions, thus shifting the activation frequencies from valence to conduction interband transitions to higher energies with higher separation between them. This results in a slower change of slope in the optical conductivity and, as the pressure increases, the optical conductivity arising from valence to conduction interband transitions is closer to a linear behaviour $\sigma \propto \omega$, as shown in Fig.~\ref{fig:opt2}~(c). 

Since we have placed the chemical potential close to the Weyl node, the activation frequencies of the interband transitions between the valence bands are shifted as the pressure increases proportionally to the chemical potential. This translates into larger regions where the optical conductivity behaves linearly due to the increased separation between activation frequencies (see Fig.~\ref{fig:opt2}~(d)).

\section{Feasibility for light dark matter detection}

\begin{table}[t]
\centering
	\begin{tabular}{c|c c c}
		& 1\% & 2\% & 3\% \\\hline
		$v_F^s$ & 0.397 & 0.339 & 0.250 \\
		$v_F^a$ & 0.552 & 0.507 & 0.423 \\
		$v_F$ & 0.945 & 0.956 & 0.882 \\
		$\delta$ & 0.896 & 1.409 & 1.640 \\
	\end{tabular}
\caption{Fermi velocities of Table~\ref{tab:fitvals} in units of $10^5$m/s.}
\label{tab:fitvals2}
\end{table}

For a material with a linear dispersion to be a realistic candidate to detect light dark matter ($m$eV deposition energies) it is necessary to fulfill the following main criteria: i) small gap ($m$eV); ii) small Fermi velocity; iii) small photon screening at energies close to the energy deposition range; and iv) scalable material growth. As we now explain, our results indicate that Ag$_3$AuSe$_2$ meets the first three criteria. The extent to which they are met varies with the lattice constant and suggesting that realizing a tunable detector is possible. 

Regarding point i), if one is to detect dark matter with $k$eV mass, it is necessary that the band gap is of the order of the deposition energy $E_D\sim m$eV. Additionally the detector should be kept at a temperature lower than this energy scale to reduce undesired thermal noise. As we have discussed, Ag$_3$AuSe$_2$ with different lattice constants, achieved by different growth rates or hydrostatic pressures, can reach the $m$eV range, satisfying point i).

The scattering between dark matter and the target material is kinematically constrained if the target velocity, the Fermi velocity, is faster than the largest possible dark matter velocity~\cite{DMfito},  $v_{\mathrm{max}} \sim 2.6 \times 10^{-3}c$. From Fig.~\ref{fig:Bands} it is already apparent that the valence bands close to the $\Gamma$ point are relatively flat. More precisely, from Table \ref{tab:fitvals2} we see that all Fermi velocities are $v_F \sim 10^{-4}c < v_{\mathrm{max}}$, hence satisfying point ii).

The real part of the optical conductivity determines the absorption of the material. As described in \cite{DMfito,Griffin18} for the scattering amplitude between the incident dark matter and the target to be large, it is beneficial that the photon is not strongly screened by the medium where it propagates, compared to a metal. This requirement is satisfied for a narrow gap Dirac material because the real part of the optical conductivity scales linearly with frequency~\cite{DMfito,Carbotte16} and the imaginary part of the dielectric tensor $\epsilon = 1 +i \sigma/\omega$ remains small as a function of frequency. For Ag$_3$AuSe$_2$ at $3\%$ (see Fig. \ref{fig:opt2} (c)) the optical conductivity grows linear as well, resulting in a dielectric constant that renders a small in-medium polarization for the photon, a necessary condition for a large scattering rate, as listed in point iii). 

Finally, a dark matter detector must be sensible to a small number of counts per year, for which the target material must be grown as large and pure as possible. Currently we are not aware of whether it is possible to grow Ag$_3$AuSe$_2$ in large crystals and defect free, yet we expect that our results should encourage experimental efforts in this direction.

The discussion regarding points i)-iii) above indicate that growing Ag$_3$AuSe$_2$ with different lattice constants can screen different, and possibly overlapping regions of parameter space. Although a full analysis of the detection capabilities is out of the scope of this work, the above discussion suggests that a detector that combines different samples of Ag$_3$AuSe$_2$ with different lattice parameters can be useful to screen different ranges of dark matter masses.

\section{Conclusion}

In this work we have studied the band structure of chiral two narrow gap semiconductors, Ag$_3$AuSe$_2$ and Ag$_3$AuTe$_2$, as well as the optical conductivity for different values of lattice parameters of the most promising dark matter detector candidate, Ag$_3$AuSe$_2$. We showed that it satisfies three important requirements to be a candidate as a target material for light dark matter detection: a $m$eV gap, shallow Fermi velocities, and small absorption. Our work sets the basis for an in depth study of the capabilities of Ag$_3$AuSe$_2$ as a light dark matter detector along the lines of Ref.~\cite{DMfito} that can consider finite momentum scattering and map the precise phase space accessible to a detector based on this material. 

As a final outlook, we note that the absence of mirror symmetries in these materials allows them to present other interesting optical responses that are only allowed in chiral space groups. These include the gyrotropic magnetic effect, which is the rotation of the polarization plane of light as it transverses the material, and a quantized circular photogalvanic effect, which is a photocurrent that grows linear in time induced by circularly polarized light set  by  fundamental  constants  only~\cite{Flicker18}.

In conclusion, we expect that our results wil encourage the growth of pure and large Ag$_3$AuSe$_2$ crystals that can serve to study chiral optical phenomena and may help to design scalable and tunable dark matter detectors in the future, based on the changes in the band structure caused by different lattice constants.

\section{Acknowledgments}
We acknowledge support from the European Union's Horizon 2020 research and innovation programme under the Marie-Sklodowska-Curie grant agreement No. 754303 and the GreQuE Cofund programme (M. A. S. M). A. G. G. is also supported by the ANR under the grant ANR-18-CE30-0001-01 and the European FET-OPEN SCHINES project No. 829044. MGV  acknowledges the IS2016-75862-P national project of the Spanish MINECO. A.B. acknowledges financial support from the Spanish Ministry of Economy and Competitiveness (FIS2016-76617-P) and the Department of Education, Universities and Research of the Basque Government and the University of the Basque Country (IT756-13).\\

\bibliographystyle{unsrt}
\bibliography{Bibliography}
\end{document}